\documentclass[pra,twocolumn,aps,floatfix,showpacs,tightenlines,superscriptaddress,amsmath,amssymb,showkeys,10pt,nofootinbib]{revtex4-2}
\usepackage{slashed}
\usepackage{xcolor}
\usepackage{graphicx}
\usepackage{hyperref}
\usepackage{placeins}
\usepackage{dsfont}
\usepackage{amsthm}

\newcommand{\be}{\begin{equation}}\newcommand{\ee}{\end{equation}}
\newcommand{\bea}{\begin{eqnarray}}\newcommand{\eea}{\end{eqnarray}}
\newcommand{\brr}{\begin{array}}\newcommand{\err}{\end{array}}
\newcommand{\bit}{\begin{itemize}}\newcommand{\eit}{\end{itemize}}
\newcommand{\ben}{\begin{enumerate}}\newcommand{\een}{\end{enumerate}}

\newcommand{\bbm}{\begin{bmatrix}}\newcommand{\ebm}{\end{bmatrix}}
\newcommand{\ba}{\begin{array}}
\newcommand{\ea}{\end{array}}

\newtheorem{mydef}{Definition}
\newtheorem{Lemma}{Lemma}
\newcommand{\bd}{\begin{mydef}} \newcommand{\ed}{\end{mydef}}
\newcommand{\bthe}{\begin{theorem}} \newcommand{\ethe}{\end{theorem}}
\newcommand{\ble}{\begin{Lemma}} \newcommand{\ele}{\end{Lemma}}

\newcommand{\dr}{\mathrm{d}}

\def\lan{\langle}
\def\lf{\left}

\def\non{\nonumber}\def\ran{\rangle}

\def\ri{\right}
\def\al{\alpha}
\def\de{\delta}\def\De{\Delta}

\def\si{\sigma}

\def\1{{_{1}}}\def\2{{_{2}}}

\newcommand{\ide}{1\hspace{-1mm}{\rm I}}

\def\noHe0{:\;\!\!\;\!\!:H_e(0):\;\!\!\;\!\!:}
\def\noHm0{:\;\!\!\;\!\!:H_\mu(0):\;\!\!\;\!\!:}

\def\lan{\langle}
\def\lf{\left}

\def\non{\nonumber}
\def\ran{\rangle}

\def\ri{\right}

\def\al{\alpha}
\def\de{\delta}\def\De{\Delta}

\def\si{\sigma}

\def\1{{_{1}}}\def\2{{_{2}}}

\makeatletter
\let\save@mathaccent\mathaccent
\newcommand*\if@single[3]{%
  \setbox0\hbox{${\mathaccent"0362{#1}}^H$}%
  \setbox2\hbox{${\mathaccent"0362{\kern0pt#1}}^H$}%
  \ifdim\ht0=\ht2 #3\else #2\fi
  }
\newcommand*\rel@kern[1]{\kern#1\dimexpr\macc@kerna}
\newcommand*\widebar[1]{\@ifnextchar^{{\wide@bar{#1}{0}}}{\wide@bar{#1}{1}}}
\newcommand*\wide@bar[2]{\if@single{#1}{\wide@bar@{#1}{#2}{1}}{\wide@bar@{#1}{#2}{2}}}
\newcommand*\wide@bar@[3]{%
  \begingroup
  \def\mathaccent##1##2{%
    \let\mathaccent\save@mathaccent
    \if#32 \let\macc@nucleus\first@char \fi
    \setbox\z@\hbox{$\macc@style{\macc@nucleus}_{}$}%
    \setbox\tw@\hbox{$\macc@style{\macc@nucleus}{}_{}$}%
    \dimen@\wd\tw@
    \advance\dimen@-\wd\z@
    \divide\dimen@ 3
    \@tempdima\wd\tw@
    \advance\@tempdima-\scriptspace
    \divide\@tempdima 10
    \advance\dimen@-\@tempdima
    \ifdim\dimen@>\z@ \dimen@0pt\fi
    \rel@kern{0.6}\kern-\dimen@
    \if#31
      \overline{\rel@kern{-0.6}\kern\dimen@\macc@nucleus\rel@kern{0.4}\kern\dimen@}%
      \advance\dimen@0.4\dimexpr\macc@kerna
      \let\final@kern#2%
      \ifdim\dimen@<\z@ \let\final@kern1\fi
      \if\final@kern1 \kern-\dimen@\fi
    \else
      \overline{\rel@kern{-0.6}\kern\dimen@#1}%
    \fi
  }%
  \macc@depth\@ne
  \let\math@bgroup\@empty \let\math@egroup\macc@set@skewchar
  \mathsurround\z@ \frozen@everymath{\mathgroup\macc@group\relax}%
  \macc@set@skewchar\relax
  \let\mathaccentV\macc@nested@a
  \if#31
    \macc@nested@a\relax111{#1}%
  \else
    \def\gobble@till@marker##1\endmarker{}%
    \futurelet\first@char\gobble@till@marker#1\endmarker
    \ifcat\noexpand\first@char A\else
      \def\first@char{}%
    \fi
    \macc@nested@a\relax111{\first@char}%
  \fi
  \endgroup
}
\makeatother
\begin{document}

\title{No-signaling-in-time condition for three-flavor neutrino oscillations}

\author{Massimo Blasone}
\email{blasone@sa.infn.it}
\affiliation{Dipartimento di Fisica, Universit\`a degli Studi di Salerno, Via Giovanni Paolo II, 132 I-84084 Fisciano (SA), Italy}
\affiliation{INFN Gruppo Collegato di Salerno - Sezione di Napoli c/o Dipartimento di Fisica, Universit\`a di Salerno, Italy}

\author{Fabrizio Illuminati}
\email{filluminati@unisa.it}
\affiliation{INFN Gruppo Collegato di Salerno - Sezione di Napoli c/o Dipartimento di Fisica, Universit\`a di Salerno, Italy}
\affiliation{Institute of Nanotechnology of the National Research Council of Italy, CNR-NANOTEC, Lecce Central Unit, c/o Campus Ecotekne, Via Monteroni, 73100 Lecce, Italy}
\affiliation{Dipartimento di Ingegneria Industriale, Universit\`a degli Studi di Salerno, Via Giovanni Paolo II, 132 I-84084 Fisciano (SA), Italy}

\author{Luciano Petruzziello}
\email{lupetruzziello@unisa.it}
\affiliation{Institut f\"ur Theoretische Physik, Albert-Einstein-Allee 11, Universit\"at Ulm, 89069 Ulm, Germany}

\author{Kyrylo Simonov}
\email{kyrylo.simonov@univie.ac.at}
\affiliation{Fakult\"{a}t f\"{u}r Mathematik, Universit\"{a}t Wien, Oskar-Morgenstern-Platz 1, 1090 Vienna, Austria}

\author{Luca Smaldone}
\email{lsmaldone@unisa.it}
\affiliation{Dipartimento di Fisica, Universit\`a degli Studi di Salerno, Via Giovanni Paolo II, 132 I-84084 Fisciano (SA), Italy}
\affiliation{INFN Gruppo Collegato di Salerno - Sezione di Napoli c/o Dipartimento di Fisica, Universit\`a di Salerno, Italy}

\begin{abstract}
Macrorealism is a fundamental concept in quantum foundations, capturing basic features of our everyday experience. Among the main tools introduced to test violations of macrorealism are the Leggett--Garg inequalities. These inequalities provide necessary and sufficient conditions for a certain data set including the correlators measured in a number of different experimental runs (weak version of macrorealism). Leggett--Garg inequalities have proved useful for detecting violations of macrorealism in neutrino oscillation experiments. However, it is interesting to investigate a set of necessary and sufficient conditions for strong macrorealism -- no-signaling-in-time and arrow-of-time conditions -- in order to obtain a more complete characterization of macrorealism in the context of neutrino oscillations. In particular, here we extend previous study of the no-signaling-in-time condition to the case of three-flavor neutrino oscillations, including the CP-violating phase of the Pontecorvo--Maki--Nakagawa--Sakata matrix.
\end{abstract}

\maketitle

\section{Introduction}

The concept of macroscopic realism (\textit{macrorealism}) was introduced in 1985 by Leggett and Garg in order to encode the intuition we develop from our everyday experience of macroscopic world, in conflict with predictions of quantum mechanics \cite{PhysRevLett.54.857}. Far from being only an abstract speculation, such an idea can be quantitatively tested, thus determining the quantum nature of a phenomenon. The most popular tests of macrorealism are related to the \emph{Leggett--Garg inequalities (LGIs)} \cite{PhysRevLett.54.857, Huelga1996, Leggett2002, Ruskov2006, Brukner2007, Brukner2008, PalaciosLaloy2010, Knee2012, PhysRep2014, Kumari2017, Zhang2025, Rybotycki2026}: violations of LGIs indicate a violation of macrorealism, giving a benchmark of quantumness of the underlying system. 

LGIs are usually viewed as the temporal counterparts of Bell inequalities \cite{Bell1964, Clauser1969, FreedmanClauser1972, ClauserHorne1974, Aspect1982, Fine1982}, however, the parallel is not perfect as spelled out in \cite{Clemente2016, Halliwell2017, Halliwell2019}. Suitable combinations of LGIs suffice for a weaker implementation of macrorealism, where different two-time correlations are measured in different experiments \cite{halliwellpla, Halliwell2017, Halliwell2019}. The same is valid for other variants of LGIs, such as their \emph{Wigner form} (WLGIs) \cite{Saha2015}. Nevertheless, it is still possible to find a set of equalities known as \emph{no-signaling-in-time} (NSIT) \cite{Brukner2008, Clemente2015} and \emph{arrow-of-time} (AoT) conditions that offer a one-to-one correspondence with the macrorealism under the former, more stringent requirements \cite{Kofler2013, Clemente2015}. Indeed, the quantumness of neutrino oscillations can be unveiled by NSIT/AoT conditions even under the fulfillment of LGIs, e.g., when wave-packet decoherence is considered \cite{Blasone:2023iwf}.

The interplay between high-energy physics and the study of quantum foundations has become one of the most active and stimulating areas of contemporary physics research \cite{Bertlmann2003, Blasone2008, Blasone2009, Donadi2012, Donadi2013, Bahrami2013, Blasone2014, SimonovLetter2016, SimonovPaper2016, Formaggio2016, Gango2017, Dixit:2018gjc,Song:2018bma, Naikoo:2019gme, Naikoo:2020LGI, Ming:2020nyc, Wang:2020vdm, Zhang2021, Dixit2021, Petruzziello2021, Li:2021fft, Jha:2022yik, Santini2022, Halliwell2022, Wang2022, Blasone:2023mbc, Blasone:2023iwf, Dixit:2023fke, Wang:2023rbf, Bosso2023, Chattopadhyay:2023xwr, Ettefaghi:2023zsh, Bouri:2024kcl, Blasone:2024dud, Blasone:2024jzv, Blasone:2024jur, Jha:2025ekn, ElBouzaidi:2025mwh, Yadav:2026mnw, Dixit:2026yft, Banerjee:2025gau, Benerjee:2026, Banerjee:2024lih, Loulijat:2026a}. A particularly compelling example of this fruitful connection is provided by the experimental observation of LGI violations in neutrino oscillation experiments \cite{Formaggio2016, Wang_2017, Wang2022}. This result has motivated extensive theoretical investigations of macrorealism in both neutrino and meson oscillation systems \cite{Naikoo:2019gme, NaikooThree2019, Shafaq:2021lju, Blasone:2023mbc, Blasone:2023iwf, Blasone:2024jur, Konwar:2024nwc, Vardani:2025vdu}. In particular, Ref. \cite{Blasone:2023iwf} analyzed the NSIT/AoT conditions for two-flavor neutrino oscillations, confirming that these conditions provide a more informative test of macrorealism than the LGIs and WLGIs. Indeed, they are able to reveal violations of macrorealism even in regimes where the LGIs are no longer violated, such as at large propagation times, when neutrino wave-packet decoherence occurs.

This paper extends the analysis of Ref. \cite{Blasone:2023iwf} investigating the NSIT and AoT conditions in the context of three-flavor neutrino oscillations, explicitly incorporating the CP-violating phase of the Pontecorvo--Maki--Nakagawa--Sakata (PMNS) mixing matrix. As in our previous studies of two-flavor neutrinos and meson oscillations, the NSIT/AoT conditions reduce to a single non-trivial equality, which is violated by neutrino oscillations. This condition is then compared with the LGIs and the WLGIs.
 
The remainder of the paper is organized as follows: in Section~\ref{sec:2}, we review the notion of macrorealism and the related conditions we will employ in our analysis (i.e. LGIs, WLGIs and NSIT/AoT). Then, in Section~\ref{neutSec}, we briefly review the basics of neutrino oscillations theory. In Section~\ref{neutNSITSec} we  compute NSIT for neutrino oscillations and compare the results with the ones on LGIs and WLGIs, while Section~\ref{conclusions} is devoted to conclusions.

\section{Tests of Macrorealism}\label{sec:2}

Quantum mechanics provides an accurate description of nature, supported by an enormous amount of experimental evidence. Nevertheless, our everyday experience appears to suggest a breakdown of genuinely quantum features, such as the superposition principle, and is usually well described by the laws of classical physics. This naturally raises the question of how the conceptual separation between these two “worlds” should be understood. To address the quantum-to-classical transition, several interpretations and modifications of quantum theory have been proposed, including Bohmian mechanics~\cite{Bohm1952I,Bohm1952II} and spontaneous collapse models~\cite{GRW1986,GPR1990,Bassi2013}. One possible way to investigate this transition is to probe coherence at macroscopic scales, which requires suitable device-independent witnesses of the validity of our intuitive picture of the everyday macroscopic world. An important step in this direction is the notion of macrorealism introduced in Ref. \cite{PhysRevLett.54.857}:

\begin{mydef}\label{def:macrorealism}
    A macrorealistic system $S$ satisfies the two following assumptions:
    \begin{itemize}
\item
\textbf{Macrorealism per se}: given a set of available macroscopically distinct states, $S$ is in one of them at any given time;
\item
\textbf{Non-invasive measurability}: it is possible in principle to determine the state of $S$ without affecting its subsequent dynamical evolution\footnote{Although usually taken for granted, a further assumption known as \textbf{Induction} can be imposed, according to which future measurements cannot affect the present state of $S$ \cite{Halliwell2017, Vitagliano2023}.}.
\end{itemize}
\end{mydef}
\noindent
In close analogy with local realism and Bell inequalities, macrorealism can be formulated in terms of quantitative conditions that are experimentally testable and, in principle, falsifiable. However, in contrast to local realism, macrorealism admits several inequivalent formulations, depending on how the requirement of non-invasive measurability in Definition \ref{def:macrorealism} is implemented. Experimental tests of macrorealism are based on the statistics of $N$ dichotomic measurements of a macroscopic observable $\mathbb{O}$, performed at times $t_0,t_1,\ldots,t_{N-1}$ on a physical system $S$. Each measurement yields a random outcome $O_i := O(t_i)\in\{-1,1\}$. For simplicity, in the following, we restrict ourselves to $N=3$.

Consider an experimental test in which $\mathbb{O}$ is measured two times in each run. Repeating the experiment for the different pairs of measurement times yields the data set consisting of three mean values
\begin{equation}\label{eq:mean}
    \langle O_i \rangle = \sum_{O_i \in \{ -1, 1\}} O_i P(O_i),
\end{equation}
where $i \in \{0,1,2\}$, and the three two-time correlation functions
\begin{equation}\label{eq:corr}
    C_{ij} := \langle O_i O_j \rangle = \sum_{O_i, O_j \in \{ -1, 1\}} O_i O_j P(O_i, O_j),
\end{equation}
where $ij \in \{01,02,12\}$, and $P(O_i)$ and $P(O_i, O_j)$ denote the corresponding single-time and two-time probability distributions. Requiring this set of quantities to satisfy non-invasive measurability leads to the weak notion of macrorealism. In this case, macrorealism holds if and only if \eqref{eq:mean}--\eqref{eq:corr} satisfy a set of inequalities constraining the corresponding temporal correlations, namely the three-time LGIs:
\bea
\label{lgi0121}
  \mathcal{L}_1(t_0, t_1, t_2) = 1+ C_{01} + C_{12} + C_{02} \geq 0 \, , \\[2mm]
\label{lgi0122}
  \mathcal{L}_2(t_0, t_1, t_2) = 1- C_{01} - C_{12} + C_{02} \geq 0 \, , \\[2mm]
  \label{lgi0123}
  \mathcal{L}_3(t_0, t_1, t_2) = 1+ C_{01} - C_{12} - C_{02} \geq 0 \, , \\[2mm]
   \label{lgi0124}
  \mathcal{L}_4(t_0, t_1, t_2) = 1- C_{01} + C_{12} - C_{02} \geq 0 \, ,
\eea
and two-time LGIs:
\bea
\label{lgit0121}
  \tilde{\mathcal{L}}_1(t_i, t_j) = 1 + \langle O_i \rangle + \langle O_j \rangle + C_{ij} \geq 0 \, , \\[2mm]
\label{lgit0122}
  \tilde{\mathcal{L}}_2(t_i, t_j) = 1 - \langle O_i \rangle - \langle O_j \rangle + C_{ij} \geq 0 \, , \\[2mm]
  \label{lgit0123}
  \tilde{\mathcal{L}}_3(t_i, t_j) = 1 + \langle O_i \rangle - \langle O_j \rangle - C_{ij} \geq 0 \, , \\[2mm]
   \label{lgit0124}
  \tilde{\mathcal{L}}_4(t_i, t_j) = 1 - \langle O_i \rangle + \langle O_j \rangle - C_{ij} \geq 0 \, ,
\eea
where $ij \in \{01, 02, 12\}$ \cite{Halliwell2017, Halliwell2019}. 
Observing a violation of the LGIs~\eqref{lgi0121}--\eqref{lgi0124} signals a breakdown of macrorealism in the system under investigation. Such a violation may therefore be interpreted as a witness of the system's ``quantumness'', in the sense that the outcome statistics predicted by quantum mechanics cannot be reproduced by any macrorealistic model satisfying these inequalities.

The above LGIs are not the only conditions that can be derived from the assumptions of macrorealism. Taking the data set as the two-time probabilities, in close analogy with Bell inequalities in Wigner form, alternative necessary constraints on the outcome statistics have been proposed in the literature, directly bounding the joint probabilities $P(O_i,O_j)$ rather than the corresponding correlation functions, known as WLGIs \cite{Saha2015,Kumari2017, Naikoo:2019gme}
\begin{align} \label{wlgi}
        \non \mathcal{W}_1(t_0, t_1, t_2) & = P(O_1,O_2)-P(-O_0,O_1)-P(O_0,O_2) \\[2mm]
        &\leq 0 , \\[2mm]
        \non \mathcal{W}_2(t_0, t_1, t_2) & = P(O_0,O_2)-P(O_0,-O_1)-P(O_1,O_2) \\[2mm]
        &\leq 0 , \label{wlgi2} \\[2mm]
        \non \mathcal{W}_3(t_0, t_1, t_2) & = P(O_0,O_1)-P(O_1,-O_2)-P(O_0,O_2) \\[2mm]
        &\leq 0 , \label{wlgi3}
    \end{align}
where $-O_i$ is the measurement outcome alternative to $O_i$, i.e., $-O_i = \mp 1$ if $O_i = \pm 1$. The underlying idea is that macrorealistic outcome statistics entails the existence of a global joint probability distribution for the outcomes at all times $t_0,\ldots,t_{N-1}$. Thus, for instance, in the considered case $N=3$, the two-time probability distributions $P(O_i,O_j)$ should be obtainable by marginalizing the global distribution $P(O_0,O_1,O_2)$. Starting from the positivity condition $P(O_0,O_1,O_2)\geq 0$, one obtains the WLGIs.

A strong reading of non-invasive measurability requires that the outcome statistics of $\mathbb{O}$ is unaffected by how many sequential measurements are performed within a single experimental run. In this case, the data set is larger than the set \eqref{eq:mean}--\eqref{eq:corr} and consists of one-, two- and three-time probabilities $P(O_i)$, $P(O_i, O_j)$, $P(O_i, O_j, O_k)$ with $i,j,k \in \{0,1,2\}$, i.e., seven independent quantities. This version of macrorealism can be characterized through two sets of \textit{equalities} constraining this data set: the NSIT and AoT conditions~\cite{Clemente2015}, given by
\bea \label{nsit1}
&& \mathrm{NSIT}^{(1)}: \ \ P(O_2) \ = \ \sum_{O_1} P(O_1, O_2) \, , \\[2mm] \label{nsit2}
&& \mathrm{NSIT}^{(2)}: \ \ P(O_0, O_2) \ = \ \sum_{O_1} P(O_0, O_1, O_2) \, , \\[2mm]
&& \mathrm{NSIT}^{(3)}: \ \ P(O_1, O_2) \ = \ \sum_{O_0} P(O_0, O_1, O_2) \, , \label{nsit3} \\[2mm]
\label{AoT1} && \mathrm{AoT}^{(1)}: \ \ P(O_0, O_1) \ = \ \sum_{O_2} P(O_0, O_1, O_2) \, , \\[2mm]
\label{AoT2} && \mathrm{AoT}^{(2)}: \ \ P(O_0) \ = \ \sum_{O_1} P(O_0, O_1) \, , \\[2mm]
\label{AoT3} && \mathrm{AoT}^{(3)}: \ \ P(O_1) \ = \ \sum_{O_2} P(O_1, O_2) \, .
\eea

Accordingly, the NSIT/AoT conditions \eqref{nsit1}-\eqref{AoT3} imply both the LGIs \eqref{lgi0121}--\eqref{lgi0124} and the WLGIs \eqref{wlgi}--\eqref{wlgi3}, whereas the converse implication does not hold. This hierarchy has been explicitly shown in physically relevant systems, including neutrino oscillations in Ref.~\cite{Blasone:2023iwf} and meson oscillations in Ref.~\cite{Blasone:2024jur}. In the following, we apply these tests to three-flavor neutrino oscillations, comparing LGIs, WLGIs, and NSIT/AoT conditions and referring to both strong and weak reading of non-invasive measurability. 


\section{Neutrino oscillations: basic facts} 
\label{neutSec}

Neutrinos participate in weak-interaction processes of production and detection as flavor states, which do not coincide with the eigenstates of the kinetic Hamiltonian, namely the mass eigenstates. Although a fully consistent description of neutrino mixing is still debated within quantum field theory \cite{Beuthe:2001rc,giunti2007fundamentals,Blasone:2025iew}, it is widely accepted that, in the relativistic regime, flavor neutrino states can be described as linear superpositions of neutrino mass eigenstates \cite{Bilenky:1978nj,Bilenky:1987ty}
\be
|\nu_\si(t) \ran \ = \ \sum_{j} \, U^*_{\si j} |\nu_j(t)\ran \, , 
\ee
where the mass eigenstates evolve according to 
\begin{align}
|\nu_j(t)\ran  &=  e^{-i E_j t} |\nu_j(0)\ran, \\
\non E_j &= \sqrt{p^2+m_j^2} \\
&\approx E + \frac{m_j^2}{2E} \, .
\end{align}
The matrix $U$ is named \emph{mixing matrix}.
In the above equations, the masses $m_j$ are taken to be much smaller than the momentum and $E = p$ is the energy of a massless neutrino. 

In the case of three flavors, the mixing matrix is known as the PMNS matrix \cite{10.1143/PTP.28.870} and it can be parametrized as
\begin{widetext}
\bea \label{pmns}
U \, =\lf(\begin{array}{ccc}
  c_{12}c_{13} & s_{12}c_{13} & s_{13}e^{-i\delta} \\
  -s_{12}c_{23}-c_{12}s_{23}s_{13}e^{i\delta} &
  c_{12}c_{23}-s_{12}s_{23}s_{13}e^{i\delta} & s_{23}c_{13} \\
  s_{12}s_{23}-c_{12}c_{23}s_{13}e^{i\delta} &
  -c_{12}s_{23}-s_{12}c_{23}s_{13}e^{i\delta} & c_{23}c_{13}
\end{array}\ri) \, , 
 \eea    
\end{widetext}
with $c_{ij}=\cos\theta_{ij},  s_{ij}=\sin\theta_{ij}$, $i, j=1,2,3$ and $\de$ is the $CP$-violating phase.

If a neutrino is produced in a weak process at time $t = 0$ with a given flavor, it evolves into a superposition of flavor states at $t>0$ as
\be
|\nu_\rho(t) \ran \ = \ \sum_\al \, \lf(\sum_{j} \,U_{\al j} U^*_{\rho j} e^{-i E_j t}\ri) |\nu_\al(0)\ran \, , 
\ee
The probability amplitude to observe a neutrino with flavor $\rho$ at time $t$ if a neutrino with flavor $\si$ was produced at $t=0$ is then
\be
{\cal A}_{\si \to \rho }(t) \ \equiv \  \lan \nu_\rho(0)|\nu_\si(t)\ran\ = \ \sum_{j} \,U^*_{\si j} U_{\rho j} e^{-i E_j t} \, .
\ee
This flavor-changing phenomenon is known as \emph{flavor oscillation}. The corresponding probability is
\bea \non
P_{\si \rightarrow \rho}(t) & = & \lf|{\cal A}_{\si \to \rho }(t)\ri|^2 \\
 & = & \sum_{j,k} U_{\rho j} U_{\si k} U^*_{\rho k} U^*_{\si j} \exp\lf(-i \frac{\Delta m^2_{j k}}{2 E} t\ri), 
\eea
where $\Delta m^2_{j k} \equiv m_j^2-m_k^2$.

The expressions above are computed in the plane-wave approximation. A more realistic description is provided by the wave-packet approach, in which each mass eigenstate is dressed with a Gaussian wave packet of spatial width $\sigma_x$~\cite{Giunti1991,giunti2007fundamentals}. Following the notation of Ref.~\cite{Blasone:2023iwf}, the state produced with flavor $\si$ reads
\be
|\nu_\si(x,t)\ran \ = \ \sum_j U^*_{\si j}\,\psi_j(t,x)\,|\nu_j\ran \, ,
\ee
where the wave packets are chosen as Gaussian functions
\begin{equation*}
    \psi_j(t,x) = \lf(\sqrt{2\pi}\sigma_x\ri)^{-\frac12} e^{i\lf(px - E_j t\ri)} \exp\!\lf[-\frac{(x - v_j t)^2}{4\sigma_x^2}\ri],
\end{equation*}
with $p$ the average momentum, $\sigma_x$ the spatial spreading and $v_j = p/E_j \simeq 1 - m_j^2/(2E^2)$ the group velocity of the $j$-th packet. Projecting onto the detected flavor $\rho$, the transition amplitude reads
\begin{align}
    {\cal A}_{\si \to \rho}(x,t) &= \lan \nu_\rho(0) | \nu_\si(x,t) \ran \non \\[2mm]
    &= \sum_j U^*_{\si j} U_{\rho j}\,\psi_j(t,x) \non \\[2mm]
    &= \lf(\sqrt{2\pi}\,\sigma_x\ri)^{-\frac12} \sum_j U^*_{\si j} U_{\rho j} e^{\,i\lf(p\,x - E_j t\ri)}\non \\  
    &\qquad \cdot \exp\!\lf[-\frac{(x - v_j t)^2}{4\sigma_x^2}\ri] \, .
\end{align}
The transition probability follows by integrating the squared modulus over the spatial coordinate, $P_{\si\to\rho}(t) = \int \dr x\,|{\cal A}_{\si\to\rho}(x,t)|^2$. Carrying out the Gaussian integration of the overlap between the $j$-th and $k$-th packets, one obtains~\cite{Blasone:2023iwf}
\begin{align}
    \nonumber P_{\si \to \rho}(t) &= \ \sum_{j,k} U_{\rho j} U_{\si k} U^*_{\rho k} U^*_{\si j} \\
    & \cdot \exp\!\lf[ -i\,\frac{\Delta m^2_{jk}}{2E}\,t
- \lf(\frac{\Delta m^2_{jk}\,t}{4\sqrt{2}\,E^2 \sigma_x}\ri)^{2} \ri] \, ,
\end{align}
where the second term in the exponent is the Gaussian decoherence factor, which suppresses the interference between the $j$-th and $k$-th mass eigenstates once the propagation time exceeds the coherence time $t^{\rm coh}_{jk} = 4\sqrt{2}\,E^2 \sigma_x/|\Delta m^2_{jk}|$.

In Ref. \cite{Blasone:2023iwf} we have analyzed NSIT/AoT conditions, in the two-flavor case. The oscillation formula in that case can be obtained putting $ \theta_{13}=\theta_{23}=0$ so that the mixing matrix reduces to
\bea \label{pmns2flav}
U \, =\lf(\begin{array}{ccc}
  c_{12} & s_{12} & 0 \\
  -s_{12} &
  c_{12} & 0 \\
 0 &
  0 & 1
\end{array}\ri) \, .
 \eea
The oscillation formula thus becomes
\begin{align}
    \nonumber P_{\si \to \rho}(t) &= \frac{\sin^2(2 \theta_{12})}{2} \\
    &\cdot \lf(1-e^{-\lf(\frac{t}{t^{\rm coh}_{12}}\ri)^2 } \cos\lf(\frac{\De m_{12}^2}{2 E} t\ri)\ri), \label{2flavosc} \quad \si \neq \rho .
\end{align}

In the next section, we investigate the NSIT/AoT conditions in the context of three-flavor neutrino oscillations and compare them with the corresponding LGIs and WLGIs. Moreover we will present the comparison with the 2-flavor case of Ref. \cite{Blasone:2023iwf}.

\section{LGIs and NSIT for three-flavor neutrino oscillations} \label{neutNSITSec}

In order to derive LGIs and NSIT/AoT in the case of three-flavor neutrino oscillations, we have to choose a dichotomic observable. In our case we choose it so that it gives $+1$ for electron neutrinos and $-1$ otherwise
\bea \non
    \hat{O}^{\nu_e} &=& 2 \Pi_{\nu_e} - \ide \\[2mm] \label{eq:obs3}
    &=& |\nu_e\rangle\langle\nu_e| - \Bigl(\sum_{\si=\mu, \tau} |\nu_\si\rangle\langle \nu_\si|\Bigr).
\eea
Operationally, this observable represents an ideal dichotomic flavor-tagging measurement. The outcome $+1$ corresponds to the identification of an electron neutrino through the charged lepton produced in a charged-current interaction, whereas the outcome $-1$ groups together the muon- and tau-flavor outcomes. Thus, the measurement may be viewed as a coarse graining of the ideal flavor measurement $\{\Pi_{\nu_e},\Pi_{\nu_\mu},\Pi_{\nu_\tau} \}$ into $\{\Pi_{\nu_e},\Pi_{\nu_\mu}+\Pi_{\nu_\tau} \}$. Let us remark that to measure a neutrino flavor means to identify the related charged lepton. Furthermore, the selection of this initial condition does not affect the qualitative conclusions of our work, as different choices of observables and states only lead to different numerical parameters to interpolate, without introducing any significant discrepancy at the interpretative level.

The next ingredients are the joint probabilities. We can assume as initial condition that the neutrino is muonic, so that the joint probability can be computed as
\be\label{retro}
P(O_1,O_2) \ = \ |\!| \Pi^{O_2} \,  \hat{U}(\De t_2) \,  \, \Pi^{O_1} \, \hat{U}(\De t_1) |\nu_\mu \ran |\!|^2 \, , 
\ee
where $\hat{U}$ is the time evolution operator and $\De t_1 \equiv t_1-t_0$, $\De t_2 \equiv t_2-t_1$. $\Pi^{O_j}$ can be either $\Pi_{\nu_e} $ or $\ide-\Pi_{\nu_e} $, so we have $O_j=e$ or $O_j=\neg e$ (not $e$), respectively. 

Proceeding in this way, we find
\begin{align}
     P(e, e) &= P_{\mu \to e}(\Delta t_1) P_{e \to e}(\Delta t_2) \, , \\
    P(e, \neg e) &= P_{\mu \to e}(\Delta t_1) \Bigl( 1-P_{e \rightarrow e}(\Delta t_2)\Bigr) \, , \\
  P(\neg e, e) &= \Bigl|\sum_{\si=\mu,\tau}\mathcal{A}_{\mu \to \si} (\Delta t_1) \mathcal{A}_{\si \to e} (\Delta t_2) \Bigr|^2 \,, \\
    \nonumber P(\neg e, \neg e) &= 1 - P_{\mu \to e}(\Delta t_1) \\
    &\qquad - \Bigl|\sum_{\si=\mu,\tau}\mathcal{A}_{\mu \to \si} (\Delta t_1) \mathcal{A}_{\si \to e} (\Delta t_2) \Bigr|^2 .
\end{align}

\noindent 
Now we set $t_0 = 0$, $t_1 = t$, and $t_2 = 2t$, so that $\Delta t_1 = \Delta t_2 = t$. Moreover we explore the case $O_0 = \neg e$, $O_1 = e$, $O_2 = e$. 

\begin{figure}
\includegraphics[width=\columnwidth]{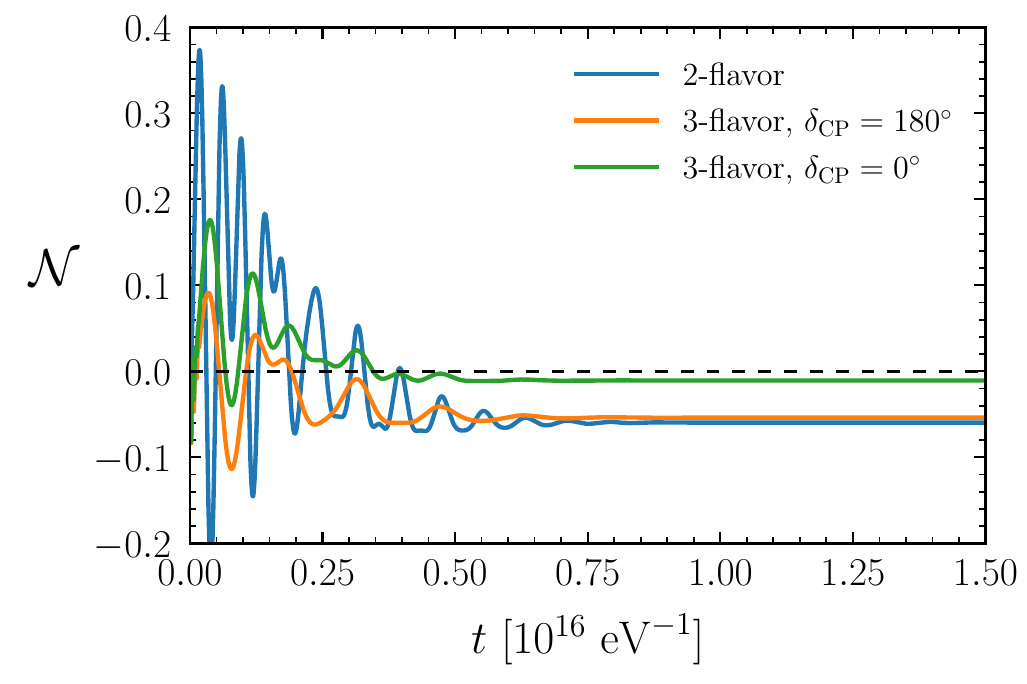}
    \caption{Comparison of NSIT condition for two-flavor oscillations and three-flavor oscillations without and with CP violation. We fix $E = 10\,\mathrm{GeV}$, $\sigma_x = 0.5\,\mathrm{GeV}^{-1}$, $\Delta m_{21}^2 = 7.92 \cdot 10^{-5}\,\mathrm{eV}^2$, $\sin^2\theta_{12} = 0.314$, and, additionally for three-flavor oscillations, $\sin^2\theta_{13} = 0.02225$, $\sin^2\theta_{23} = 0.573$, $\Delta m_{31}^2 = 2.507 \cdot 10^{-3}\,\mathrm{eV}^2$. We are adopting natural units with $\hbar=c=1$. Taking into account that meutrinos are ultrarelativistic, the distance in Km can be obtained multiplying the x-axis values by $1.97327 \times 10^{6}$.}
		\label{fignsit}
\end{figure}

We begin by analyzing the AoT and NSIT conditions that fully characterize macrorealism under strong reading of non-invasive measurability. Since AoT conditions are trivially satisfied because Eq.~\eqref{retro} cannot lead to retrocausality, the NSIT conditions reduce to a single expression
\begin{equation}
    P(O_2) = \sum_{O_1} P(O_1, O_2).
\end{equation}
Indeed, since
\begin{align*}
    P(O_2) &= P_{\mu \to e}(2t) \, , \\[2mm]
    P(e, O_2) &= P_{\mu \to e}(t) P_{e\to e}(t) \, , \\[2mm]
    P(\neg e, O_2) &= \Bigl| \mathcal{A}_{\mu \to \mu}(t) \mathcal{A}_{\mu \to e}(t) + \mathcal{A}_{\mu \to \tau}(t) \mathcal{A}_{\tau \to e}(t) \Bigr|^2 \, ,
\end{align*}
 the NSIT condition is $\mathcal{N}(t)=0$, with
\begin{align} \non
    \mathcal{N}(t) &\equiv   P(O_2)- [P(e, O_2) +P(\neg e, O_2)] \\[2mm] \label{NSIT}
			\non &= P_{\mu \to e}(2t)  - P_{\mu \to e}(t) P_{e\to e}(t) \\[2mm]
            &\quad - \Bigl| \mathcal{A}_{\mu \to \mu}(t) \mathcal{A}_{\mu \to e}(t) + \mathcal{A}_{\mu \to \tau}(t) \mathcal{A}_{\tau \to e}(t) \Bigr|^2.
\end{align}
This function encodes the effect of an intermediate measurement. The last piece represents the interference between the stories where one measures a $\tau$ and a $\mu$ neutrino (let us remind that $\Pi_{\neg e}=\Pi_{\mu}+\Pi_{\tau}$) and thus it cannot be interpreted in terms of probability.

In Fig. \ref{fignsit}, this function is plotted for the present three-flavor case and compared with the corresponding two-flavor function studied in Ref. \cite{Blasone:2023iwf} (see Eq. \eqref{2flavosc}). Let us remark that here and in the other plots we have used the current best fit for the parameters of the PMNS matrix, while we have chosen $\si_x$ to give a good illustration of our conceptual results. We find that, in both cases, the NSIT condition is violated at large times, even beyond the coherence time scale. The violation of NSIT appears to be larger in the two-flavor case, while for three-flavor case it depends on $\de_{CP}$. However, as discussed below, this feature is not confirmed by all the conditions analyzed in this work when performing a similar comparison between the two- and three-flavor cases.

\begin{figure}
\includegraphics[width=\columnwidth]{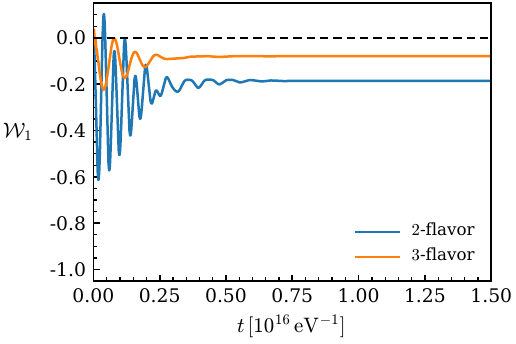}
    \caption{Comparison of Wigner LGI $\mathcal{W}_1$ condition for two-flavor and three-flavor oscillations. Other parameters are fixed as in Fig. \ref{fignsit}.}
		\label{figwlg1}
\end{figure}

\begin{figure}
\includegraphics[width=\columnwidth]{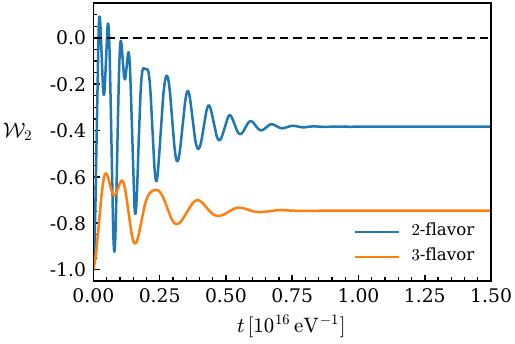}
    \caption{Comparison of Wigner LGI $\mathcal{W}_2$ condition for two-flavor and three-flavor oscillations. Other parameters are fixed as in Fig. \ref{fignsit}.}
		\label{figwlg2}
\end{figure}

Differently from the NSIT condition, for the given choice of parameter values and initial state, the Wigner LGIs for three-flavor oscillations are formally similar to the two-flavor counterpart
\begin{align*}
    \mathcal{W}_1(t) &= P_{\mu \to e}(t) P_{e \to e}(t) - P_{\mu \to e}(2t) \leq 0 \, ,  \\[2mm]
    \mathcal{W}_2(t) &=  P_{\mu \to e}(2t) -  1 + P_{\mu \to e}(t) - P_{\mu \to e}(t) P_{e \to e}(t) \leq 0 \, , \\[2mm]
    \mathcal{W}_3(t) &= \mathcal{W}_1(t) \leq 0 \, ,
\end{align*}
and only the expressions of probabilities themselves differ. The comparison between the two- and the three-flavor case is presented in Figs. \ref{figwlg1} and \ref{figwlg2}.

\begin{figure}
\includegraphics[width=\columnwidth]{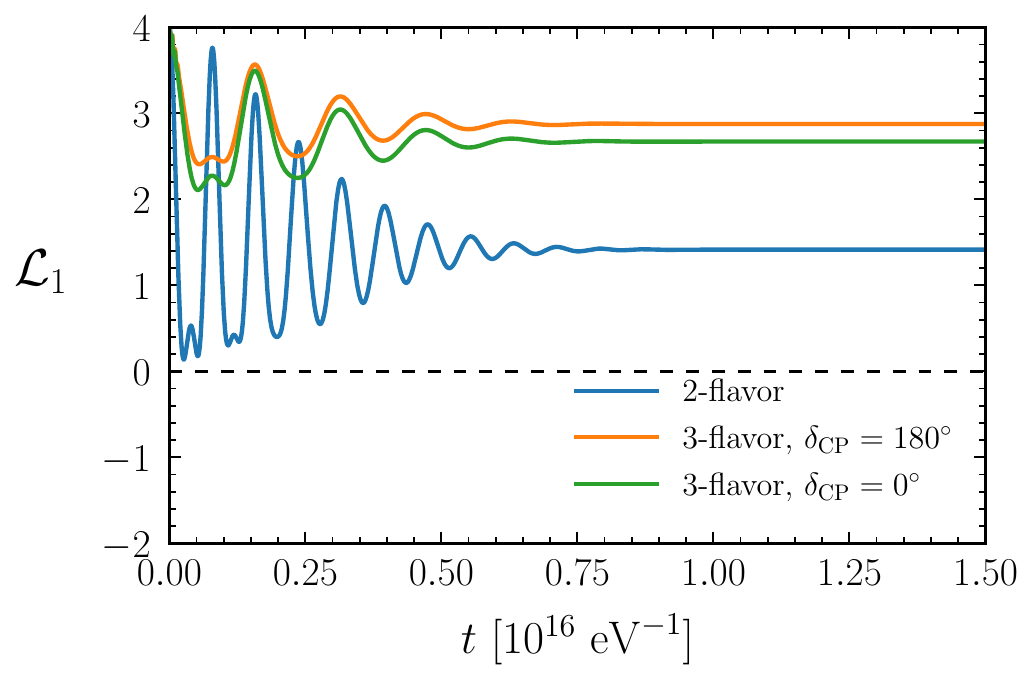}
    \caption{Comparison of standard LGI $\mathcal{L}_1$ condition for two-flavor oscillations and three-flavor oscillations with and without CP violation. Other parameters are fixed as in Fig. \ref{fignsit}.}
		\label{figlg1}
\end{figure}
\begin{figure}
\includegraphics[width=\columnwidth]{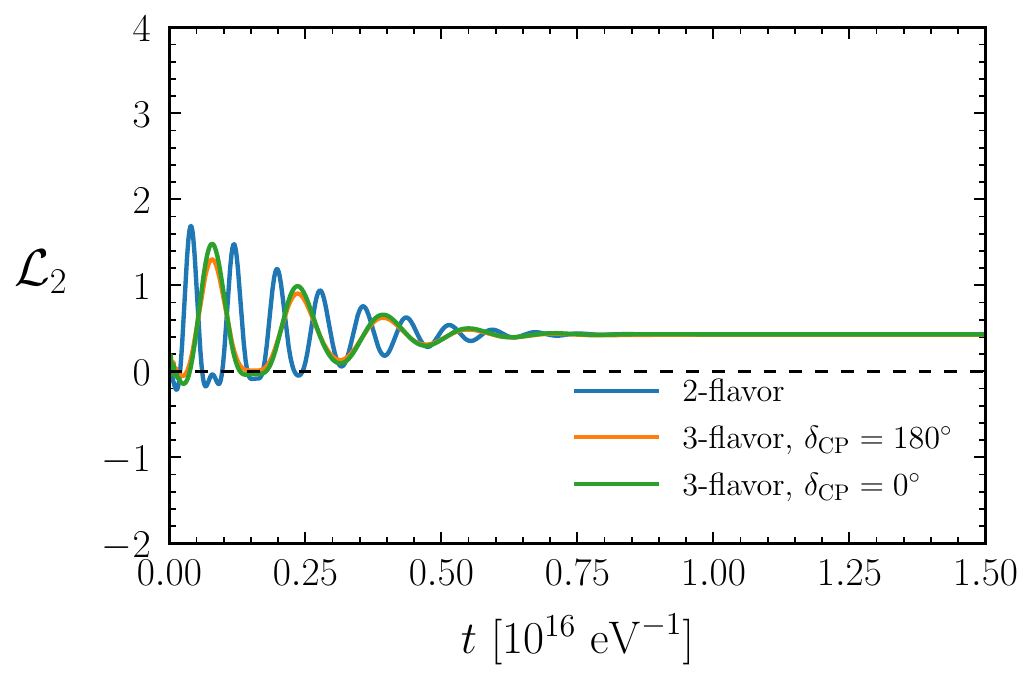}
    \caption{Comparison of standard LGI $\mathcal{L}_2$ condition for two-flavor oscillations and three-flavor oscillations with and without CP violation. Other parameters are fixed as in Fig. \ref{fignsit}.}
		\label{figlg2}
\end{figure}

\begin{figure}
\includegraphics[width=\columnwidth]{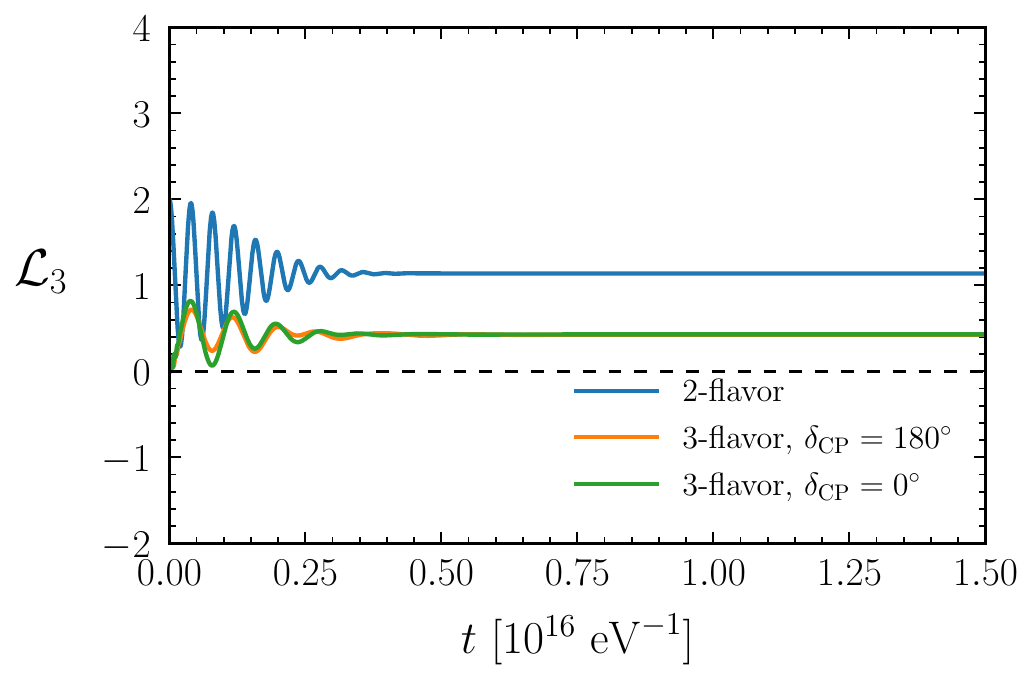}
    \caption{Comparison of standard LGI $\mathcal{L}_3$ condition for two-flavor oscillations and three-flavor oscillations with and without CP violation. Other parameters are fixed as in Fig. \ref{fignsit}.}
		\label{figlg3}
\end{figure}

\begin{figure}
\includegraphics[width=\columnwidth]{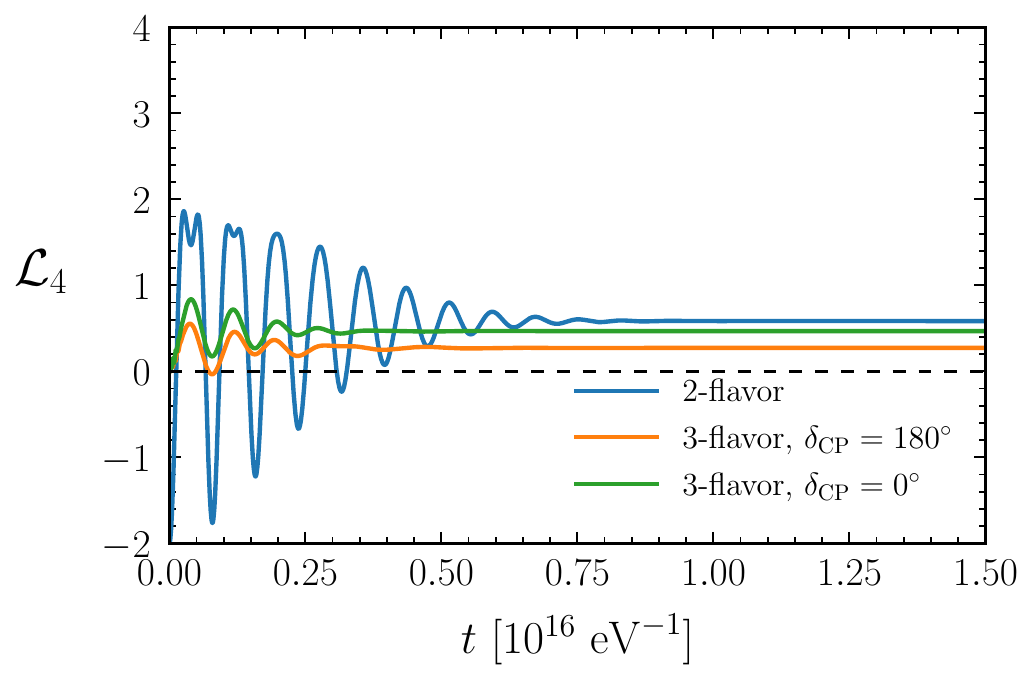}
    \caption{Comparison of standard LGI $\mathcal{L}_4$ condition for two-flavor oscillations and three-flavor oscillations with and without CP violation. Other parameters are fixed as in Fig. \ref{fignsit}.}
		\label{figlg4}
\end{figure}

In contrast to the Wigner LGIs, standard LGIs depend explicitly on $\tau$-flavor, that is,
\begin{align}
    \nonumber \label{eq:neutrinoLGI1} \mathcal{L}_1(t) &=
        4 - 4P_{\mu \to e}(t) - 2P_{\mu \to e}(2t)
        + 2P_{\mu \to e}(t)P_{e \to e}(t) \\[2mm]
    - 2 &\, \Bigl| \mathcal{A}_{\mu \to \mu}(t) \mathcal{A}_{\mu \to e}(t)
        + \mathcal{A}_{\mu \to \tau}(t) \mathcal{A}_{\tau \to e}(t) \Bigr|^2
        \geq 0 \, , \\[2mm]
    \nonumber \label{eq:neutrinoLGI2} \mathcal{L}_2(t) &=
        4P_{\mu \to e}(t) - 2P_{\mu \to e}(2t)
        - 2P_{\mu \to e}(t)P_{e \to e}(t) \\[2mm]
    + 2&\,\Bigl| \mathcal{A}_{\mu \to \mu}(t) \mathcal{A}_{\mu \to e}(t)
        + \mathcal{A}_{\mu \to \tau}(t) \mathcal{A}_{\tau \to e}(t) \Bigr|^2
        \geq 0 \, , \\[2mm]
    \nonumber \label{eq:neutrinoLGI3} \mathcal{L}_3(t) &=
        2P_{\mu \to e}(2t)
        - 2P_{\mu \to e}(t)P_{e \to e}(t) \\[2mm]
    + 2&\,\Bigl| \mathcal{A}_{\mu \to \mu}(t) \mathcal{A}_{\mu \to e}(t)
        + \mathcal{A}_{\mu \to \tau}(t) \mathcal{A}_{\tau \to e}(t) \Bigr|^2
        \geq 0 \, , \\[2mm]
    \nonumber  \label{eq:neutrinoLGI4} \mathcal{L}_4(t) &=
        2P_{\mu \to e}(2t)
        + 2P_{\mu \to e}(t)P_{e \to e}(t) \\[2mm]
    - 2 &\, \Bigl| \mathcal{A}_{\mu \to \mu}(t) \mathcal{A}_{\mu \to e}(t)
        + \mathcal{A}_{\mu \to \tau}(t) \mathcal{A}_{\tau \to e}(t) \Bigr|^2
        \geq 0 \, .
\end{align}
Importantly, while for two-flavor oscillations $\mathcal{L}_3$ and $\mathcal{L}_4$ coincide, this is not the case anymore for three-flavor oscillations. On the other hand, for the chosen measurement times, the two-time LGIs \eqref{lgit0121}--\eqref{lgit0124} coincide with the LGIs \eqref{eq:neutrinoLGI1}--\eqref{eq:neutrinoLGI4} for $ij = 12$ while being trivial for $ij \in \{01, 02\}$. Therefore, in neutrino oscillations, macrorealism under weak reading of non-invasive measuribility can be tested via the set of standard LGIs \eqref{eq:neutrinoLGI1}--\eqref{eq:neutrinoLGI4}. The comparison between the two- and the three-flavor case is presented in Figs. \ref{figlg1}, \ref{figlg2}, \ref{figlg3}, and \ref{figlg4}. In particular, as in the two-flavor case, three-flavor neutrino oscillations satisfy the standard LGIs over most of the time interval considered, except for a violation of \eqref{eq:neutrinoLGI2} at short times, which is less pronounced than in the two-flavor case.


\section{Discussion and Outlook} \label{conclusions}
In this paper, we have extended the analysis of Ref.~\cite{Blasone:2023iwf}, studying the NSIT and AoT conditions in the framework of three-flavor neutrino oscillations, explicitly taking into account the CP-violating phase of the PMNS mixing matrix. We have shown that, as in our previous analyses of two-flavor neutrinos and meson oscillations, the NSIT/AoT conditions reduce to a single non-trivial equality, which is violated in the presence of neutrino oscillations. This necessary and sufficient condition for macrorealism in its strong form has been compared with the standard LGIs, which, for the neutrino-oscillation setting considered here, characterize macrorealism in its weak form, and with their Wigner form counterparts. Notably, the inclusion of the tau flavor introduces a dependence on the CP-violating phase $\delta$ into tests of macrorealism in neutrino oscillations, affecting both the NSIT condition associated with the strong form of macrorealism and the LGIs associated with its weak form. In particular, the violation of the NSIT condition in the three-flavor case is significantly enhanced in the case $\delta \neq 0$ with respect to conserved CP symmetry. We also observe that the expression of the NSIT function Eq. \eqref{NSIT} is not entirely given in terms of oscillation probabilities, thus hinting to a potential new (quantum) observable sensitive to CP violation.

As a potential future application, we observe that the present framework may provide a novel perspective on the long-standing problem of the correct field theoretical description of neutrino mixing and  oscillations~\cite{Beuthe:2001rc, Blasone:2025iew}. Indeed, different treatments (ranging from the standard quantum-mechanical approximation to the exact flavor-vacuum formalism, which predict corrections to the oscillation formula and a non-trivial structure of the flavor states) generally lead to slightly different oscillation probabilities. Since the NSIT/AoT conditions probe the detailed temporal correlations of the system rather than time-averaged observables, they could in principle act as fine-grained witnesses able to discriminate among these competing descriptions, thereby complementing the study of LGIs in the QFT of neutrino oscillations carried out in Ref.~\cite{Blasone:2023mbc}. Assessing quantitatively whether the foreseen corrections leave a detectable imprint on the NSIT/AoT witnesses represents an interesting direction for future work.

\section*{Acknowledgements}
L. P. acknowledges networking support by the COST Action CA18108. K.~S. acknowledges: This research was funded in whole or in part by the Austrian Science Fund (FWF) 10.55776/PAT4559623.


\bibliography{main}

\bibliographystyle{apsrev4-2}

\end{document}